# The incarnation of the Nersesyan-Tsvelik model in $(NO)[Cu(NO_3)_3]$

O. Volkova[1], I. Morozov[1], V. Shutov[1], E. Lapsheva[1], P. Sindzingre[2], O. Cépas[3,4], M. Yehia[5], V. Kataev[5], R. Klingeler[5], B. Büchner[5], and A. Vasiliev[1]

[1]Moscow State University, Moscow 119991 Russia

[2]Laboratoire de physique théorique de la matière condensée, UMR7600 CNRS, Université Pierre-et-Marie-Curie, Paris 6, 75252 Paris cedex 05, France.

[3] Institut Néel, CNRS et Université Joseph Fourier, BP 166, 38042 Grenoble, France.

[4]Laboratoire J.-V. Poncelet, UMI2615 CNRS Independent University of Moscow, Moscow 119002, Russia.

[5]Leibniz Institute for Solid State and Materials Research, D-01171 Dresden, Germany

The topology of the magnetic interactions of the copper spins in the nitrosonium nitratocuprate $(NO)[Cu(NO_3)_3]$ suggests that it could be a realization of the Nersesyan-Tsvelik model, whose ground state was argued to be either a resonating valence bond (RVB) state or a valence bond crystal (VBC). The measurement of thermodynamic and resonant properties reveals a behavior inherent to low dimensional spin S = ½ systems and provides indeed no evidence for the formation of long-range magnetic order down to 1.8 K.

Low dimensional quantum magnets are currently the object of an intensive experimental and theoretical research. This stems from the rich physics which is displayed by these systems due to their reduced dimensionality and competing interactions which often push the transition to ordered states to very low temperatures or even preclude the onset of magnetism at all. Geometric frustration is one of the effects which are believed to lead to possible non-classical states. A non-classical ground state does exist in a pure one-dimensional quantum antiferromagnet which is disordered and carries low-energy spinon excitations with fractional quantum numbers. A fundamental question is whether such non classical states can survive in higher dimensions and whether they could realize the long-sought resonant valence bond (RVB) state [1]. The concept of RVB state is of utmost importance in modern condensed matter physics, not only for frustrated magnetism in general but also in the context of high – temperature superconductivity of layered cupric compounds [2].

An interesting model is the frustrated S = ½ square lattice $J_1 - J_2$ and its extensions to further neighbor interactions. Depending on the ratio between the nearest – neighbor antiferromagnetic exchange $J_1$ and the second – neighbor coupling $J_2$ this model has a Néel ground state at weak frustration, and a stripe or collinear Néel state at strong frustration. There is a narrow region between the two phases, in the range $0.4<J_2/J_1<0.6$ where there is now a consensus for the absence of magnetic order. Instead, a spin-liquid or a valence bond crystal state may be realized. The search for experimental realizations of such a model has been pursued intensively in copper and vanadium oxides (see [3] for a recent review) but the narrow region of parameters where non-classical states may appear is clearly a challenge for chemistry. An alternative is offered by a different $J_1 - J_2$ model recently introduced by Nersesyan and Tsvelik (also named the "confederate flag" model) [4]. It differs from the $J_1 - J_2$ model by the spatial anisotropy of the nearest neighbor couplings (J, J') along the horizontal and vertical directions and the same $J_2$ along the diagonals (see Fig. 1a). The model is particularly interesting for the special ratio $J'/J_2 = 2$ where the ground state was first argued to be RVB in the anisotropic limit, $J >> J' = 2J_2$ (weakly coupled chains) [4]. This result has been questioned since then and the ground state could be a VBC instead [5-8]. In any case, the special condition $J'/ J_2 = 2$ forces the effective mean-field to vanish and makes the mean-field theory of coupled chains [9] (which would predict long-range Neel order at zero temperature) inapplicable. Although it seems that this condition requires again some fine-tuning of the couplings, we shall show that it is in fact *exactly* realized in a nitrosonium nitratocuprate $(NO)[Cu(NO_3)_3]$.

The single crystals of nitrosonium nitratocuprate $(NO)[Cu(NO_3)_3]$ were obtained by means of wet chemistry according to the procedure described in Ref. [10]. The phase composition of the crystalline samples was determined by powder X-ray diffraction. The measurement was carried out on a DRON 3M difractometer using a $CuK_\alpha$ radiation in the $2\theta$ range of 5 – 60°. The single-phase nature of the obtained samples was confirmed by similarity of the experimental X-ray diffraction patterns and theoretical ones calculated from single crystal X-ray diffraction data [10]. The bluish single crystals of $(NO)[Cu(NO_3)_3]$ with dimensions (3-6)×(1.5-2.5)×(0.5-1) $mm^3$ in the form of elongated thickened plates are not stable in air and could be safely investigated in sealed glass ampoules only. The experimental investigation of $(NO)[Cu(NO_3)_3]$ consisted of measurements of magnetization by MPMS "Quantum Design" in a temperature range 1.8 – 300 K and electron spin resonance (ESR) by an X - Band "Bruker EMX-Series" spectrometer operating at a fixed frequency $\nu$ = 9.5 GHz in a temperature range 3.4 – 300 K. Besides, the specific heat was measured at low temperatures by PPMS "Quantum design" taking special measures against decomposition of the sample.

The crystal structure of $(NO)[Cu(NO_3)_3]$ is represented by weakly coupled layers whose structure is shown in Fig. 1b. Assumingly, the strongest interaction J between $Cu^{2+}$ (S = ½) ions is provided via $NO_3^-$ groups forming therefore infinite horizontal chains along the *b* - axis. These chains are coupled via $NO_3^-$ and $NO^+$ ions in *bc* plane in such a manner that vertical exchange interaction along the *c* axis, J', is *exactly* twice the exchange interaction along the diagonal, $J_2$ : there are two symmetric superexchanges paths contributing to J' whereas there is only one (and equivalent by symmetry) contributing to $J_2$. The interplane coupling along the *a* axis is assumed to be weak and unfrustrated. Two equivalent exchange interaction routes along this axis pass via $NO_3$ group but through apical oxygen at a Cu – O distance 2.539 Å. This is to be compared with single exchange interaction pass via $NO_3$ group through basal oxygen at a Cu – O distance 1.985 Å. The weakness of interplane coupling follows from magnetic inactivity of $d_{z2}$ orbital oriented along the *a* – axis. Therefore, the topology of $(NO)[Cu(NO_3)_3]$ magnetic subsystem is exactly that of the two – dimensional "confederate flag" model J >> J' = $2J_2$ (cf. Fig. 1a).

The temperature dependence of specific heat C in $(NO)[Cu(NO_3)_3]$, shown in Fig. 2, evidences upturn at low temperatures which could be somewhat suppressed by magnetic field. While the measurements were not performed down to sufficiently low temperatures the upturns in specific heat measured at H = 0 and H = 1.5 T can be treated as the shoulder of the Schottky anomaly. In this case, the data taken at H = 0 T can be fitted by a sum of linear contribution γT related to the one – dimensional antiferromagnetic magnons, cubic term $\beta T^3$ related to phonons, and Schottky contribution $C_{Sch}$. The resulting formula can be written as:

$$C = \gamma T + \beta T^3 + \alpha \frac{3}{2} R \left( \frac{\Delta}{kT} \right)^2 \frac{\exp\left( -\frac{\Delta}{kT} \right)}{\left( 1 + 3\exp\left( -\frac{\Delta}{kT} \right) \right)^2} .$$

The best fit of available data is obtained with γ = 0.54 J/molK$^2$, β = 0.0016 J/molK$^4$, α = 0.33 and Δ = 5 K. That same parameters allow good fit of specific heat measured at H = 1.5 T. The sensitivity of the Schottky anomaly to external magnetic field indicates magnetic origin of the relevant two – level system. The value of weighting factor α = 0.33 makes it difficult to ascribe it either to extrinsic or intrinsic contribution. At the same time, we mention that the shape and curvature of low temperature upturn do not suggest that the system approaches the phase transition to long range magnetically ordered state.

The temperature dependence of the magnetic susceptibility χ of $(NO)[Cu(NO_3)_3]$ taken in a magnetic field 0.1 T oriented in the *bc* plane is shown in Fig. 2. On lowering the temperature, the magnetic susceptibility χ first increases, passes through a broad maximum, and then rapidly increases again, showing a pronounced Curie-like behavior. The origin of strong low-temperature upturn is not clear since the method of preparation excludes the presence in the structure of any other cations except $Cu^{2+}$ and $NO^+$ and any other anions except $NO_3^-$ while high optical quality of the available crystals is apparent. The pronounced Curie-like behavior cannot be explained also by a Schottky type anomaly with $\Delta = 5$ K. The broad maximum in $\chi(T)$ can be seen as a signature of the low dimensionality of $(NO)[Cu(NO_3)_3]$ magnetic subsystem.

In order to study the *intrinsic* spin susceptibility and to obtain insights into the spin dynamics, we have performed ESR measurements of a single crystalline sample for two orientations of the external magnetic field: The in-plane orientation parallel to the $CuO_4$ plaquettes, and the out-of-plane orientation, perpendicular to the $CuO_4$ plaquettes, $\|$ and $\perp$, respectively. For both directions the ESR spectrum consists of a single line (field derivative of the absorption) with the shape very close to a Lorentzian. The fit of the experimental signal to the Lorentzian derivative line profile enables an accurate determination of the intensity of the ESR signal $I_{ESR}$, the peak-to-peak linewidth $\Delta H_{pp}$ and the resonance field $H_{res}$. The $g$-factor tensor calculated from the resonance field as $g = h\nu/\mu_B H_{res}$ yields the values $g_{\|} = 2.06$ and $g_{\perp} = 2.36$. Here $h$ is the Planck constant and $\mu_B$ is the Bohr magneton. The obtained g-factor values are typical for a $Cu^{2+}$ ion in planar square ligand coordination [11]. Remarkably, the $H_{res}$ and correspondingly the $g$-values practically do not depend on temperature (see Fig. 4, inset), i.e. there are no indications for the development of local internal fields due to the onset of (quasi) static short- or long-range order in the entire temperature range of the study. This strongly supports our conjecture of the low dimensionality of the spin-1/2 Heisenberg lattice in $(NO)[Cu(NO_3)_3]$ where magnetic order is not expected at a finite temperature.

Generally, the integrated intensity of the ESR signal $I_{ESR}$ is directly proportional to the static susceptibility of the spins participating in the resonance [12]. Its analysis enables therefore an insight into the intrinsic magnetic susceptibility $\chi_{spin}$ of the spin lattice in $(NO)[Cu(NO_3)_3]$. The temperature dependence of the $I_{ESR}$ normalized to its value at 295 K is shown in Fig. 4 for the $\|$ and $\perp$ magnetic field orientations. For both field directions these curves are very similar and, if compared to the static magnetic measurements, show even more pronounced low dimensional behavior.

In order to extract some information on the magnetic couplings, we have calculated the spin susceptibility by exact diagonalization of the Nersesyan-Tsvelik model (varying $\alpha = J'/J$). We have used clusters of up to 24 spins with different geometries (a square of 4x4 spins, a ladder of 8x2 spins, and a stripe of 6x4 spins with periodic boundary conditions). Because of finite-size effects, the susceptibility is exact only for $T > 2T_{max}$ where $T_{max}$ is the temperature of the broad maximum of the susceptibility $\chi_{max}$ (see Fig. 5, solid and dashed lines for $\alpha = 0.6$, for instance). It is therefore difficult to access to the low-temperature regime where the susceptibility decreases. Nonetheless, it appears that the product $\chi_{max}T_{max}$ is less sensitive to the size and is especially useful since we know the exact Bonner-Fisher result for decoupled chains ($\alpha$ = 0), given by $\chi_{max}T_{max} = 0.0941 N_A g^2 \mu_B^2/k_B$ [13]. When $\alpha \neq 0$, the product is a pure function of $\alpha$ and is shown in Fig. 4 (inset). We see that $\chi_{max}T_{max}$ is approximately linear in $\alpha$ and the slope is found to be -0.0558 (for the 8x2), -0.0614 (for the 4x4) and -0.0607 (for the 6x4), so is weakly size-dependent. We note that the result at $\alpha$ = 0 almost coincides with Bonner-Fisher for the 8x2 cluster (this is the shape with the longer chains). Combining these results, we have $\chi_{max}T_{max}/(N_A g^2 \mu_B^2/k_B)$= 0.0941-0.060$\alpha$. We now compare with the experiments, using $g^2$ = 4.68 from ESR (powder average), and the conversion to standard units gives $\chi_{max}T_{max}$ = 0.165-0.105 $\alpha$ (emu.K/mol). Given that, experimentally, $\chi_{max}T_{max}$ = 0.163±0.007 emu.K/mol, we can conclude that $-0.05 < \alpha < 0.09$. Note that to obtain a precise value of $\chi_{max}$ experimentally we have subtracted a Curie impurity tail from the susceptibility measurement and the subtraction fits well the intrinsic ESR susceptibility as shown in Fig. 3. The system is therefore in the weak coupling regime: given the error bar, the interchain couplings could be either ferromagnetic or antiferromagnetic but we can exclude a strong coupling regime. J can then be estimated from the position of the maximum of the susceptibility for infinite decoupled chains, $T_{max}$=0.6408J [14], giving J=170K.

Another important information can be obtained from the analysis of the linewidth $\Delta H_{pp}$ which particularly in spin-1/2 systems is mainly determined by the relaxation rate of the spin fluctuations perpendicular to the applied field. The $\Delta H_{pp}$ for both orientations of the external field shows remarkably strong temperature dependence. In particular, below ~ 100 K the linewidth decreases by almost an order of magnitude which at first glance might be interpreted as a strong depletion of the spin fluctuation density due to the opening of the spin gap. In fact, in this temperature regime the $\Delta H_{pp}$ ($T$) dependence can be phenomenologically reasonably well described by an exponential function ~ $e^{-\Delta/T}$ with an energy gap $\Delta_{||} \approx \Delta_{\perp} \sim 77$ K. However, the finite ESR intensity, i.e. the finite intrinsic spin susceptibility, observed down to the lowest temperature contradicts the spin gap scenario. Alternatively, it is known than in a 2D antiferromagnet at temperatures far above the magnetic ordering temperature $T_N$ the linewidth is determined mainly by the long-wave $q \approx 0$ fluctuation modes whose strength decreases with lowering the temperature as $\chi_{spin}T$ (see, e.g., [15]). If a competing contribution due to the 'short-wave' spin fluctuations at the staggered wave vector $q = \pi$ remain small, e.g. if the spin system is still in the regime $T >> T_N$, one could indeed expect even for a gapless situation a progressive strong narrowing of the ESR signal due to the reduction of both the $\chi_{spin}$ and the temperature. The product of the normalized ESR intensity, i.e. the $\chi_{spin}$, and the temperature is shown in Fig.

6. The $\chi_{spin}T$ curve is scaled to match most closely the $\Delta H_{pp}$ ($T$) curves. Indeed, one observes a reasonably good qualitative agreement between the $\chi_{spin}T$ and the $\Delta H_{pp}$ ($T$) dependence. One notices that a substantial anisotropy of the linewidth at high temperatures strongly decreases at low temperatures (Fig. 5) whereas the anisotropy of the *g*-factor stays constant (Fig. 4, inset). Such a reduction of the linewidth anisotropy is expected if the strength of $q \approx 0$ modes decreases [15].

There is also a surprising similarity between the temperature behavior of the ESR linewidth for $(NO)[Cu(NO_3)_3]$ and that for some 1D spin-1/2 systems like, e.g., $KCuF_3$ [16,17], a spin-Peierls compound $CuGeO_3$ [18], and a quarter-filled spin ladder $NaV_2O_5$ [19]. In all of them a similarly strong temperature dependence of $\Delta H_{pp}$ concomitant with a strong temperature variation of the anisotropy of the linewidth is ascribed to the Dzyaloshinskii-Moriya (DM) interaction, which is allowed by the crystal symmetry in those systems. In the present case too, one can infer from the crystal structure that the DM interaction is present between spins along the chain and the D-vector is staggered from bond to bond (because the Cu ion is at an inversion center, but is forbidden for the interchain couplings. Since the XZ plane passing through the middle of the Cu-Cu bond is a mirror plane, the D vector should be perpendicular to the Cu-Cu bond, anywhere inside that plane. An interesting consequence is the field-induced gap [20], as for Cu-benzoate, although in the present case, the gap should be small with the small fields applied, but it may affect the low temperature susceptibility as well.

It is of course difficult to assess the nature of the ground state of $(NO)[Cu(NO_3)_3]$ in the view of the present experimental results. Nersesyan and Tsvelik have argued that, for small interchain couplings, the ground state remains disordered and realizes a chiral π-flux RVB spin-liquid at zero temperature [4], which the present results do not contradict, in fact. However the situation is not yet settled: extension towards finite inter-chain couplings has lead to consider other candidates for the ground state, such as valence bond crystals [5, 6]. The claim for a VBC is not supported by a DMRG calculation for a spin ladder, though, [8] but may not be excluded for the infinite system [5]. Although the system would be gapped in this case, the value of gap $\Delta = 5$ K estimated from the specific heat measurements seems to be rather high. In fact the spin gap was claimed to be extremely small from numerical studies [9]. It is also consistent with the idea that the system could be nearly critical: there are Neel states away from the special line $J'=2J_2$ but very close to it in the parameter space $J'$-$J_2$ [5, 6]. Accordingly, the ground state of $(NO)[Cu(NO_3)_3]$ could be a gapless spin liquid. This does not contradict to results of thermodynamic and resonant measurements. However such a gapless spin liquid state could be unstable to additional interactions (unfrustrated interplane interactions, DM interactions) which can result in three–dimensional long–range magnetic order at lower temperatures.

In conclusion, we have found that a nitrosonium nitratocuprate $(NO)[Cu(NO_3)_3]$ seems to be a good realization of the Nersesyan-Tsvelik model in the weak coupling regime: the main magnetic couplings of the Heisenberg model were found to be J = 170 K and $-0.05 < \alpha = J'/J < 0.09$. It is clear that a precise determination of the interchain couplings deserves more studies. In addition, departures from the Heisenberg model in the form of Dzyaloshinskii-Moriya interactions certainly exist (as suggested by ESR) and may also contribute to the low-temperature susceptibility. In any case, thanks to the special geometry of the Nersesyan-Tsvelik model, the interchain interactions are not only weak but also strongly frustrated, thus making possible to realize an RVB or VBC state. Experimentally, indeed, no indications for the phase transition were found down to 1.8 K despite strong antiferromagnetic couplings, and it is an interesting issue as to whether such states are realized or not in the present material.

Figure captions

Fig. 1. The schematic representation of anisotropic confederate flag model (a), the schematic representation of crystal structure of $(NO)[Cu(NO_3)_3]$. Note, J' = $2J_2$. Green spheres are the $Cu^{2+}$ ions. The dumbbells represent the $NO^+$ cations groups. The $NO_3^-$ anions groups are represented by tilted and flat triangles. Note, J' = $2J_2$ (b).

Fig. 2. The temperature dependences of specific heat C measured at H = 0 T (o) and H = 1.5 T (□) in $(NO)[Cu(NO_3)_3]$. The fitting curves are shown with solid lines.

Fig. 3. The temperature dependences of magnetic susceptibility taken at H = 0.1 T (o) and normalized magnetic susceptibility from ESR data (□) of $(NO)[Cu(NO_3)_3]$. The fitting curves with Curie – Weiss law and sum of Curie – Weiss term with normalized magnetic susceptibility from ESR are shown with dashed and solid lines.

Fig. 4. The temperature dependence of the intensity of ESR spectra $I_{ESR}$ normalized to the room temperature value for a magnetic field parallel and perpendicular to the plane of $CuO_4$ plaquettes. Inset: the temperature dependence of the values of the *g*–factor tensor $g_{||}$ and $g_{\perp}$.

Fig 5. Susceptibility from exact diagonalizations (α=J'/J characterises the interchain coupling). N is the system size and various geometries are used [4x4, 8x2, 6x4]. Inset: $\chi_{max}T_{max}/(N_a g^2 \mu_B^2/k_B)$ as a function of α. The exact result for decoupled chains (α=0) is shown by a square.

Fig 6. Temperature dependence of the ESR linewidth $\Delta H_{pp}$ for a magnetic field parallel and perpendicular to the plane of $CuO_4$ plaquettes (symbols). Dash line is the scaled product $\chi_{spin}(T)\cdot T$.

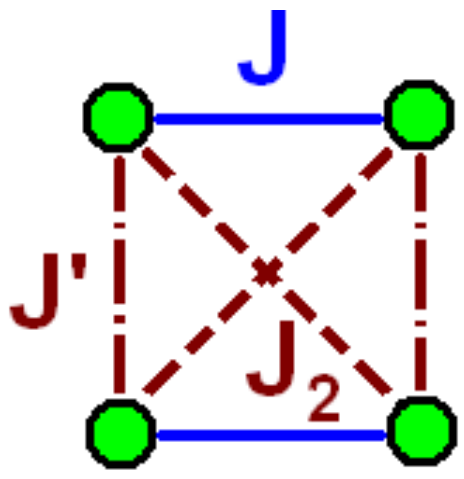

Fig. 1 a.

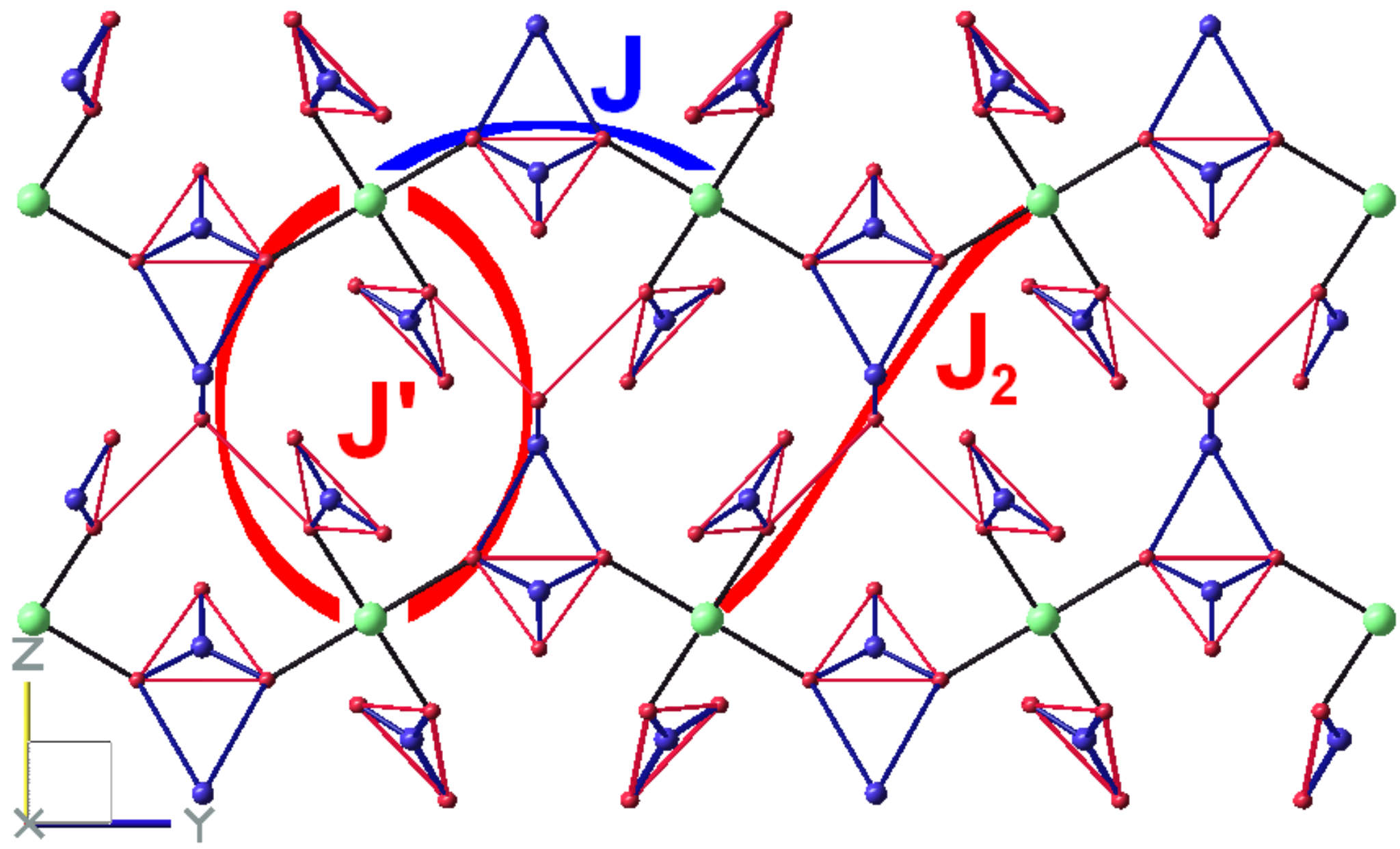

Fig. 1 b.

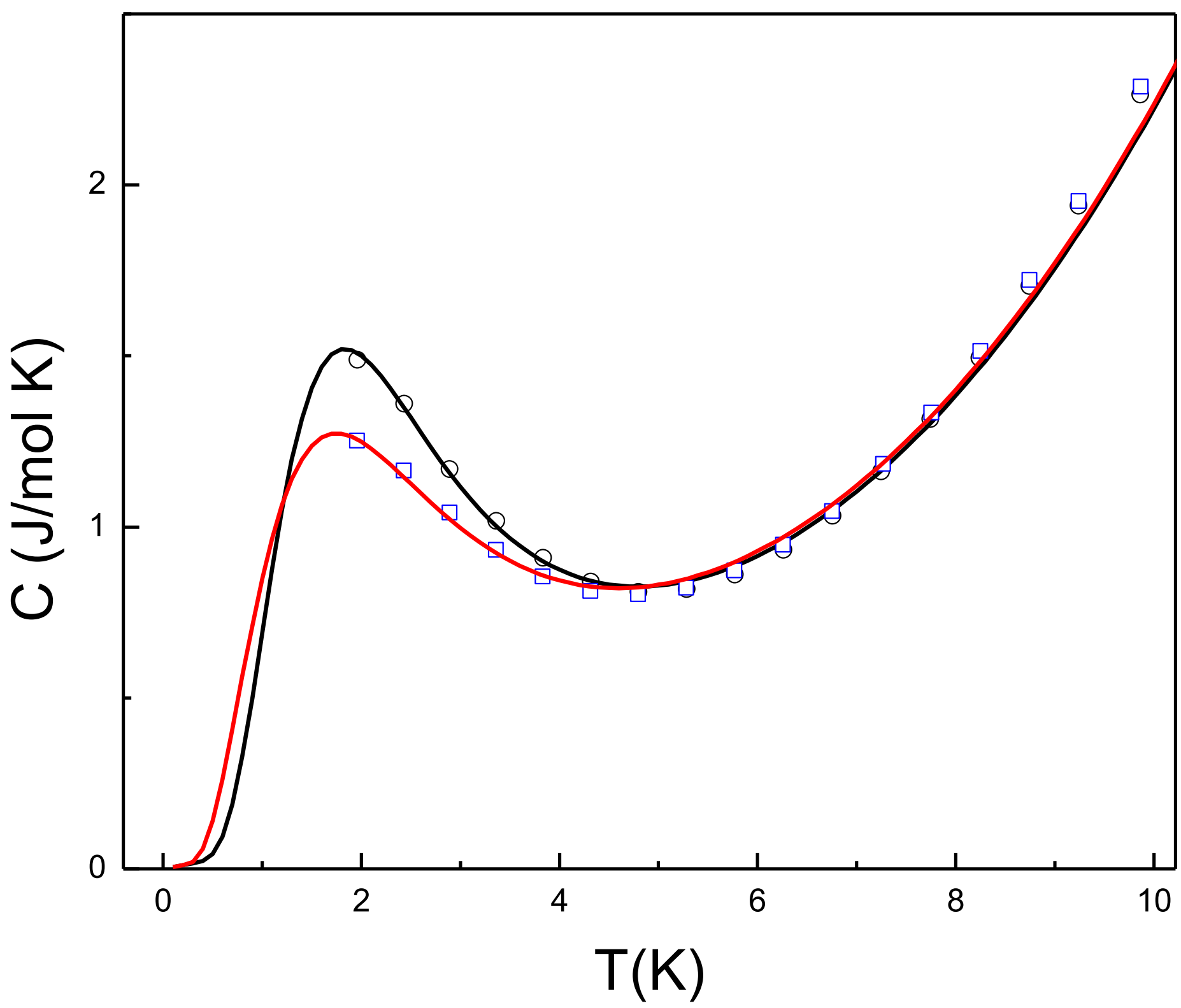

Fig. 2.

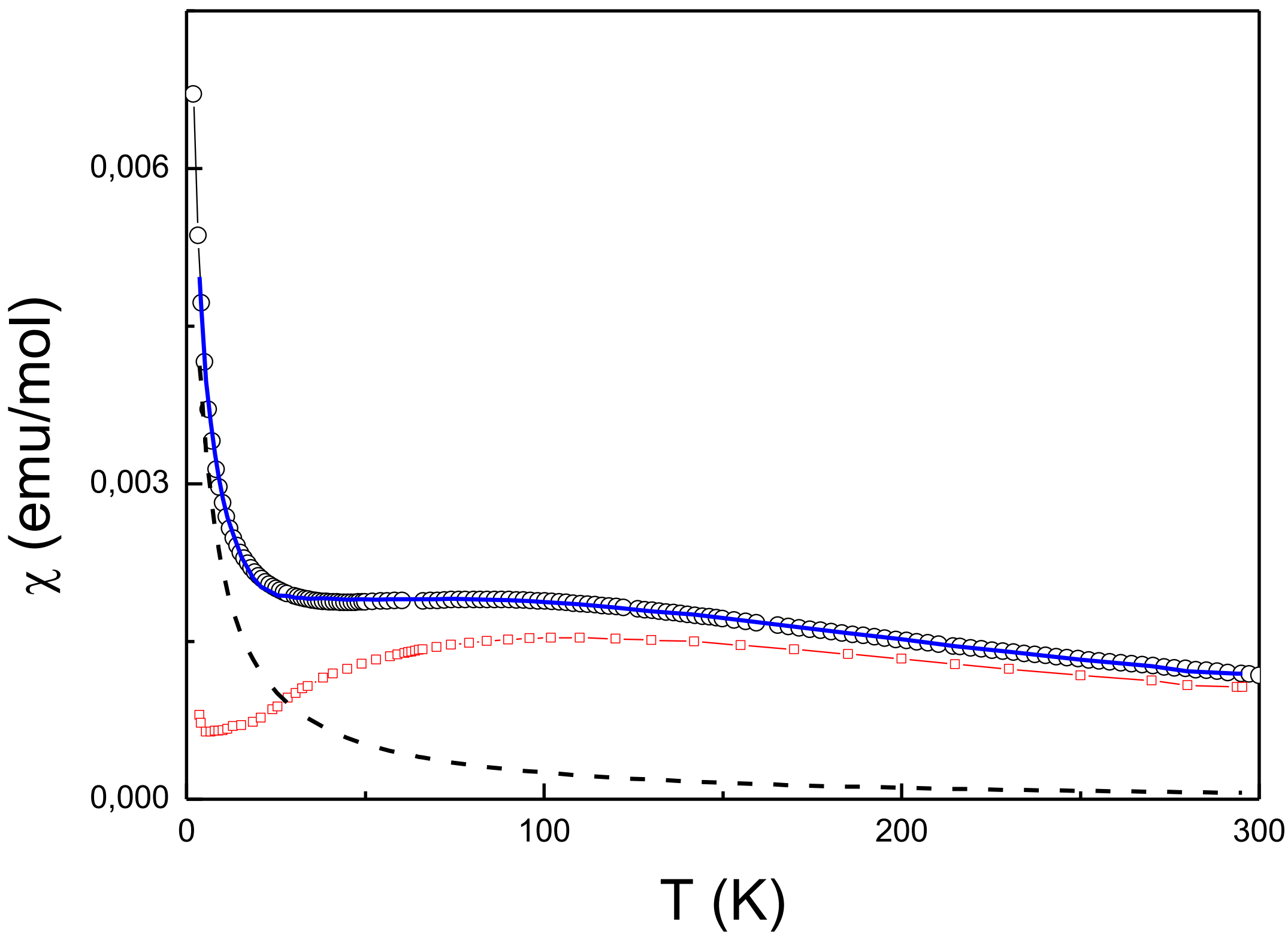

χ (emu/mol)
0,006
0,003
0,000
0
100
200
300
T (K)

Fig. 3.

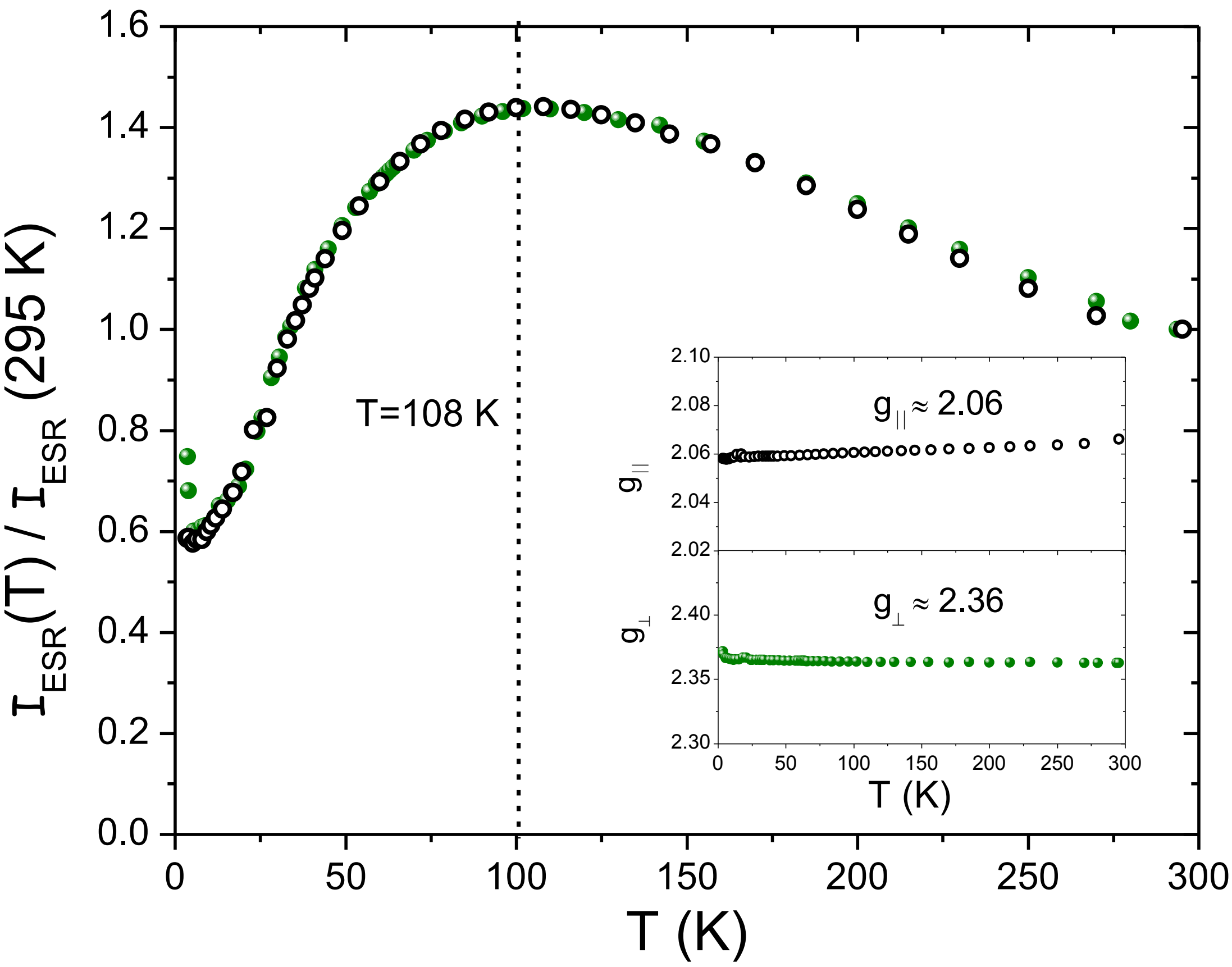

$I_{ESR}(T) / I_{ESR}(295\ K)$
1.6
1.4
1.2
1.0
0.8
0.6
0.4
0.2
0.0
0
50
100
150
200
250
300
T (K)
T=108 K
$g_{||} \approx 2.06$
$g_{\perp} \approx 2.36$
$g_{||}$
$g_{\perp}$
2.10
2.08
2.06
2.04
2.02
2.40
2.35
2.30
T (K)

Fig. 4.

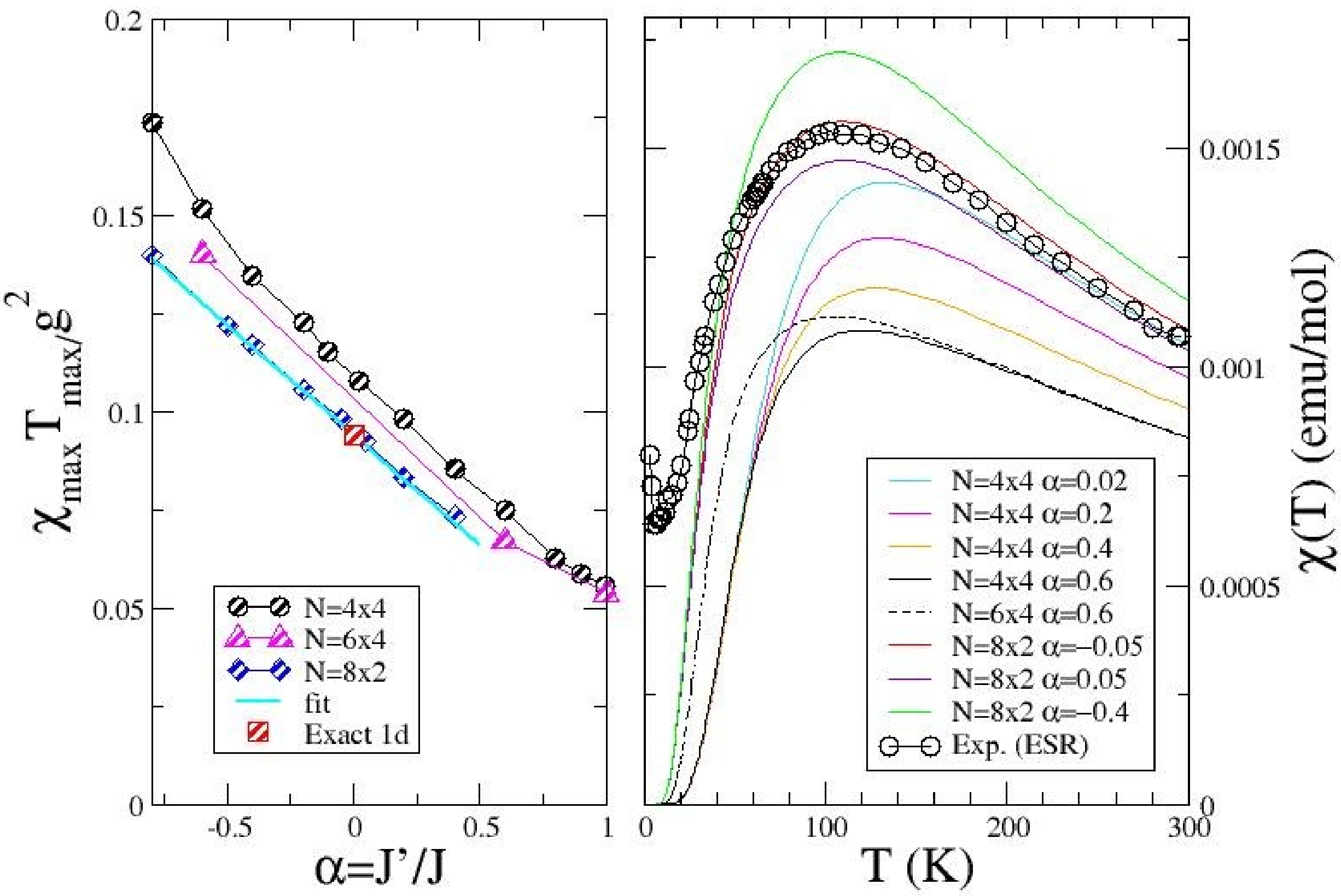

0.2
0.15
0.1
0.05
0
-0.5
0
0.5
1
$\chi_{max} T_{max}/g^2$
$\alpha$=J'/J
N=4x4
N=6x4
N=8x2
fit
Exact 1d
0
100
200
300
T (K)
0.0015
0.001
0.0005
0
χ(T) (emu/mol)
N=4x4 α=0.02
N=4x4 α=0.2
N=4x4 α=0.4
N=4x4 α=0.6
N=6x4 α=0.6
N=8x2 α=−0.05
N=8x2 α=0.05
N=8x2 α=−0.4
Exp. (ESR)

Fig. 5.

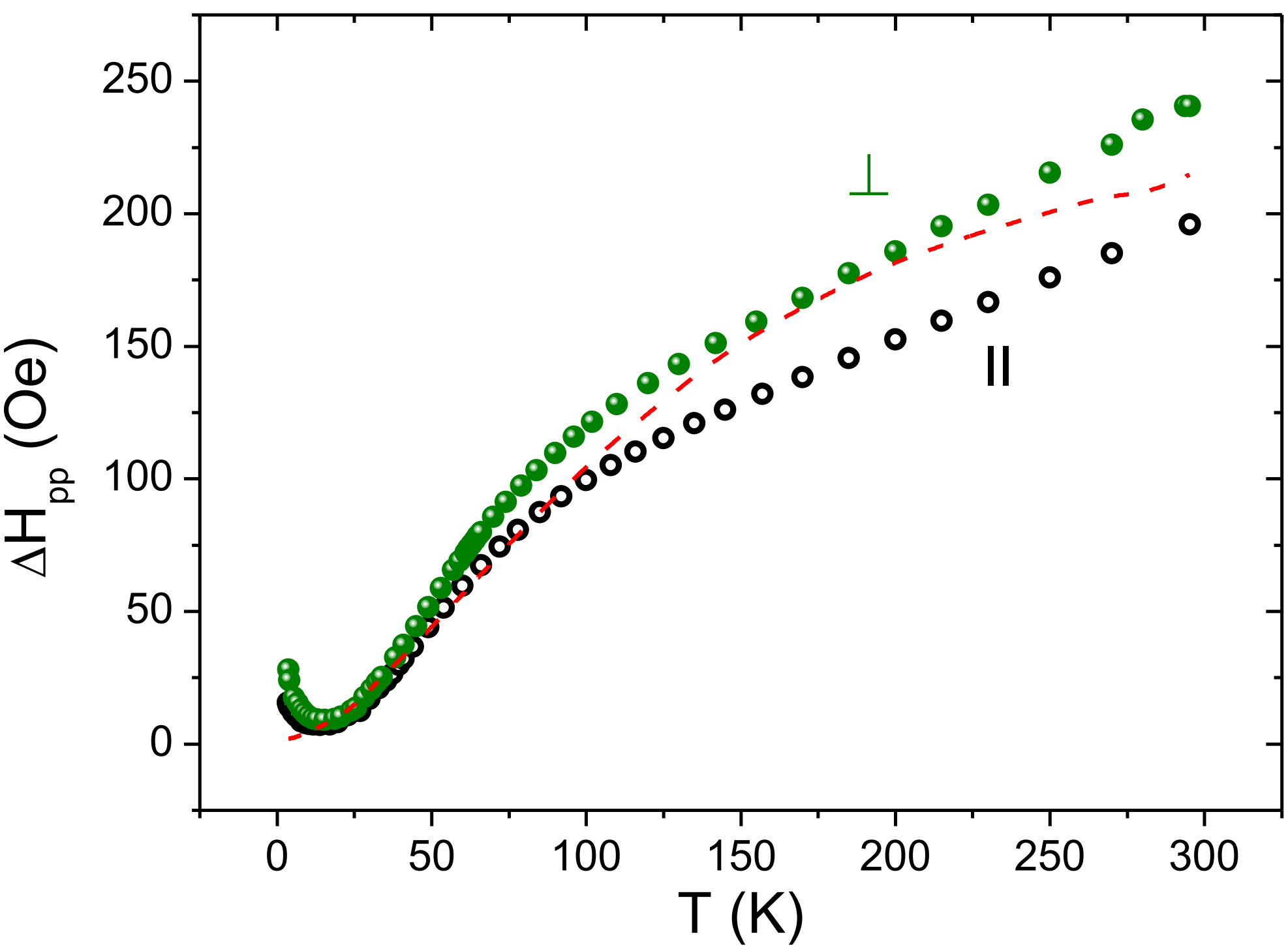

250
200
150
100
50
0
$\Delta H_{pp}$ (Oe)
0
50
100
150
200
250
300
T (K)
⊥
II

Fig. 6.